\documentclass[b5paper,twoside]{jpconf}  %Default declaration%
\usepackage{graphicx}
\usepackage{picinpar}

\newcommand{\nh}{$\rm{N}_{\rm H}$ }
\newcommand{\av}{$\rm{A}_{\rm V}$ }

\begin{document}
\title[The Galactic \nh Relation]{The Galactic \nh- \av Relation and its Application to Historical Galactic SNRs}
\author[Wen-Wu TIAN, Hongquan SU, F.Y. XIANG]{Wen-Wu TIAN$^1$, Hongquan SU$^{2,1}$, F.Y. XIANG$^3$}
\address{$^1$National Astronomical Observatories, Chinese Academy of Sciences, Beijing, 100012, China, tww@bao.ac.cn}
\address{$^2$Department of Astronomy, Beijing Normal University, Beijing 100008, China, hq\_su@mail.bnu.edu.cn}
\address{$^3$Department of Physics, Xiangtan University, Xiangtan, Hunan, 411
105, China}
\begin{abstract}
We refine a classic relation between the hydrogen column density (\nh) and optical extinction (\av) by employing 39 Galactic Supernova Remnants (SNRs) with X-rays, optical and/or infra-red data available. We find \nh $=(1.69\pm0.07) \times 10^{21}$ \av cm $^{-2}mag^{-1}$. Applying this relation to three Galactic SNRs with good historical records allows us to further constrain either their progenitor's distances or magnitudes, which is independent access to their distances.
\end{abstract}
\begin{table*}
\tiny{ \begin{flushleft}
\centering
\caption{The \nh and \av of the SNRs}
\begin{tabular}{ccccc}
  \hline\hline
  SNR Name & \nh (10$^{21}$ cm$^{-2}$) & Model$^1$ & \av (mag.) & Method \\
  \hline\hline
  G4.5$+$6.8         & 5.7$\pm$0.8                & PL                & 2.2$\pm$0.6       & spectra \\
  G6.4$-$0.1         & $5.1^{+0.8}_{-0.2}$        & NEI               & 3.57$\pm$0.47     & H$_{\alpha}$/H$_{\beta}$ \\
  G11.2$-$0.3        & $21.9^{+1.4}_{-1.6}$       & NEI               & 13$\pm$1.3        & FeII Ratio  \\
  G13.3$-$1.3        & 1.0$\pm$0.1                & two temperature   & 0.47$\pm$0.05     & H$_{\alpha}$/H$_{\beta}$\\
  G34.7$-$0.4        & $8.9^{+1.5}_{-0.5}$        & NEI               & 10$\pm$1          & $-$ \\
  G39.7$-$2.0        & $5.9^{+2.3}_{-1.9}$        & PL                & 2.65$\pm$1.55     & H$_{\alpha}$/H$_{\beta}$ \\
  G43.3$-$0.2        & $5.24^{+0.10}_{-0.11}$     & Two-CIE           & 7.5$\pm$2.5       & $-$ \\
  G49.2$-$0.7        & 17.1$\pm$0.8               & PL                & 8.0$\pm$0.8       & $-$ \\
  G53.6$-$2.2        & $7.7^{+7.7}_{-6.8}$        & TP                & 3.57$\pm$0.47     & H$_{\alpha}$/H$_{\beta}$\\
  G54.1$+$0.3        & $15.4^{+2.0}_{-1.9}$       & TB                & 8.0$\pm$0.7       & $-$ \\
  G65.3$+$5.7        & 1.4$\pm$0.1                & $-$               & 0.265$\pm$0.085   & H$_{\alpha}$/H$_{\beta}$\\
  G65.7$+$1.2        & 2.3$\pm$1.7                & BB$+$PL           & 2.23$\pm$0.22     & R-D \\
  G67.7$+$1.8        & 5.15$\pm$1.95              & TP                & 5.27$\pm$0.93     & H$_{\alpha}$/H$_{\beta}$\\
  G69.0$+$2.7        & 3.0$\pm$0.1                & PL	              & 2.48$\pm$0.25	  & H$_{\alpha}$/H$_{\beta}$\\
  G74.0$-$8.5        & $0.5^{+0.4}_{-0.0}$        & CERSM	          & 0.248$\pm$0.062	  & R-D\\
  G78.2$+$2.1	     & 11$\pm$1	                  & STTE              & 4.55$\pm$0.05	  & H$_{\alpha}$/H$_{\beta}$\\
  G89.0$+$4.7        & $3.1^{+0.3}_{-0.2}$        & Spectral          & $2.0^{+0.2}_{-0.0}$ & Near IR color\\
  G94.0$+$1.0	     & 13$\pm$0.1                 & On-Off method	  & 8.8$\pm$0.6	        & EM\\
  G109.1$-$1.0       & $6.0^{+0.2}_{-1.0}$        & Spectral          & 3.08$\pm$0.64	    & H$_{\alpha}$/H$_{\beta}$\\
  G111.7$-$2.1	     & 12.5$\pm$0.3	              & BB	              & 5.0$\pm$0.4 	    & S¢ò ratio \\
  G116.9$+$0.2       & $7.5^{0.8}_{0.7}$          & Spectral          & 2.65$\pm$0.45	    & H$_{\alpha}$/H$_{\beta}$\\
  G119.5$+$10.2	     & 2.8$\pm$0.3	              & PL+Thermal	      & $1.37^{+0}_{-0.14}$	& Far-IR emission \\
  G120.1$+$1.4	     & 4.4$\pm$0.5	              & PB	              & 1.86$\pm$0.20	    & NS \\
  G130.7$+$3.1	     & 4.53$\pm$0.09	          & PL+TP	          & 1.95$\pm$0.35	    & H$_{\alpha}$/H$_{\beta}$\\
  G166.0$+$4.3	     & 2.6$\pm$0.3	              & RS	              & 1.68$\pm$0.17	    & H$_{\alpha}$/H$_{\beta}$\\
  G180.0$-$1.7	     & $2.47^{+0.15}_{0.14}$      & BB	              & 0.72$\pm$0.07	    & H$_{\alpha}$/H$_{\beta}$\\
  G184.6$-$5.8	     & 3.45$\pm$0.15	          & $-$               & 1.46$\pm$0.12	    & SII+2200${\AA}$ \\
  G189.1$+$3.0	     & 5.8$\pm$0.6	              & $-$               & 3.5$\pm$0.5	        & $-$ \\
  G205.5$+$0.5	     & 0.8$\pm$0.1	              & $-$	              & 0.67$\pm$0.07       & H$_{\alpha}$/H$_{\beta}$\\
  G260.4$-$3.4	     & 4.1$\pm$0.2	              & BB	              & 1.5$\pm$0.2	        & $-$ \\
  G263.9$-$3.3	     & 0.22$\pm$0.12	          & RS                &	0.10$\pm$0.01	    & R-D \\
  G266.2$-$1.2	     & 3.45$\pm$0.15	          & BB	              & 2.88$\pm$1.40	    & EM  \\
  G284.3$-$1.8	     & 5.0$\pm$1.7	              & PL	              & 4.4$\pm$0.4	        & $-$ \\
  G292.0$+$1.8	     & 3.17$\pm$0.15	          & PL	              & 2.0$\pm$0.2	        & $-$ \\
  G315.4$-$2.3	     & $3.2^{+0.6}_{-0.3}$        & PL	              & 1.0$\pm$0.1	        & $-$ \\
  G320.4$-$1.2	     & 8.6$\pm$0.9	              & PL	              & 4.8$\pm$0.5	        & $-$ \\
  G327.6$+$14.6	     & 0.568$\pm$0.021	          & TP+PL	          & 0.35$\pm$0.04	    & $-$ \\
  G332.4$-$0.4	     & 6.8$\pm$0.7	              & Nebula NEI	      & 4.5$\pm$0.5	        & $-$ \\
  G332.5$-$5.6	     & 0.83$\pm$0.08	          & RS                & 0.84$\pm$0.08	    & H$_{\alpha}$/H$_{\beta}$\\
  \hline
\end{tabular}
\newline
$^1$ Model abbreviations:
AE: Absorption Edge model,
BB: Blackbody,
CERSM: Combined Edge Region Spectral Model,
EM: Emission Measurement,
NEI: non-equilibrium ionization model,
NS: near stars,
PB: Pure Bremsstrahlung model,
PL: Power-Law,
R$-$D: Reddening vs. distance plot,
RS: Raymond-Smith model,
STTE: single-temperature thermal emission,
SII$+{\AA}$: average of the two values obtained from the SII emission line ratio and 2200${\AA}$ interstellar absorption,
TB Thermal bremsstrahlung
TP: Thermal Plasma,
Vmekal: thin thermal model with variable cosmic abundance.
\end{flushleft}}
\end{table*}

   The equivalent hydrogen column density ($\rm{N}_{\rm H}$) is deduced from the X-ray extinction, which is mainly caused by the heavy elements in the interstellar media. The optical extinction ($\rm{A}_{\rm V}$) is caused by grains composed of these same heavier elements. In the Galaxy, an experiential linear relation between them was found about three decades ago, though the distribution of the ISM is not uniform in different directions.
The value of  \nh/\av varies from (2.22$\pm$0.14)$\times 10^{21}$(Gorenstein 1975) to (1.79$\pm$0.03)$\times 10^{21}$ (Predehl \& Schmitt 1995).
The latest value 2.21$\pm$0.09 was given by G\"uver \& \"Ozel (2009), who utilized the data on 22 SNRs that have been observed with the X-ray observatories
and for which optical extinction and/or reddening measurements have been performed. We collect 39 SNRs with both the \nh and \av values from the literatures
(see the Table 1). We adopt R=3.1 to obtain the \av if only the E(B$-$V) is given in literatures, i.e. \av = 3.1 $\times$ E(B$-$V).
Table 1 includes a typical 10\% error on the optical extinction for remnants whose errors are not available from literatures. We fit with Origin Pro8 and
set $1/(\rm{N}_{\rm H}^{error}/ \rm{N}_{\rm H}+ \rm A_{\rm V}^{error}/ \rm{A}_{\rm V})$ as the weight. Our result is \nh=$(1.69\pm0.07)  \times 10^{21}$ \av $cm^{-2}$mag$^{-1}$. The square of correlation coefficient R-square=0.94. The confidence level is 95\%.\\
    We apply this linear relation to three well-known type Ia SNRs to get their \av from the observed \nh. The apparent magnitude m was recorded when their supernova exploded and the absolute magnitude M is obtained from the decline of the supernova B-band brightness at maximum and 15 days later\cite{4}. With these values, we calculate the distances of the SNRs and compare the results with others' which get from the different models. This may become an independent  method to get the distance of the Galactic SNRs.\\
    \\
\begin{table*}
\tiny{\begin{flushleft}
\centering
\caption{The Application to Historical Galactic SNRs}
\begin{tabular}{ccccccccccc}
  \hline\hline
SNR     & \nh           & \av.our  & \av.other & m     & M     & d.our   & d.other   & method$^1$     \\
        & $10^{21}cm^{-2}$ & mag.        & mag.          & mag.  & mag.  & kpc     & kpc       &             \\
         \hline\hline
Tycho   & 4.4$\pm$0.5    & 2.60$\pm$0.19  & 1.86$\pm$0.20  & -4.25$\pm$0.25 & -19$\pm$0.3  & 2.69$\pm$0.11 & 2.7$\pm$0.1  & HI absor.  \\
SN1006  & 0.568$\pm$0.021& 0.336$\pm$0.001& 0.35           & -7.75$\pm$1.75    & -19$\pm$0.3  & 1.52$\pm$0.14 & 2.18$\pm$0.08& SVPM     \\
Kepler  & 5.7$\pm$0.8    & 3.4$\pm$0.3    & 2.7$\pm$0.3    & -3.0$\pm$0.5     & -19$\pm$0.3  & 3.31$\pm$0.02 & 2.9$\pm$0.4  & SVPM     \\
\hline
\end{tabular}
\newline
$^1$SVPM: the Shock Velocity and the Proper Motion
\end{flushleft}}
\end{table*}
We thank supports from NSFC (011241001) and Bairen-program of the CAS. This research was funded in part by ``the Fundamental Research Funds for the Central Universities" and by Graduate School of Beijing Normal University.

\end{document}